\documentclass[12pt]{article}
\usepackage{amsfonts}

\newcommand\BEQ{\begin{equation}}
\newcommand\EEQ{\end{equation}}
\newcommand\BEQA{\begin{eqnarray}}
\newcommand\EEQA{\end{eqnarray}}

\def\hk{hyper\-K\"ah\-ler\ }
\def\Hk{Hyper\-K\"ah\-ler\ }
\def\kh{K\"ahler\ }
\def\sm{$\sigma$--model\ }
\def\mm{moment map\ }
\def\sd{$N_c \leftrightarrow N_f-N_c$}
\def\sy{sym\-plec\-tic\ }
\def\ma{mani\-fold\ }
\def\FI{Fayet--Iliopoulos\ }
\def\SW{Seiberg--Witten\ }
\def\Lie#1{{\cal L}_{#1}}
\def\mm{{\cal P}^{x}}
\def\tr{\mbox{tr}}
\def\im{\mbox{Im}}
\def\ie{{\it i.e.\ }}

\def\ZZ{\mathbb{Z}}
\def\HH{\mathbb{H}}
\def\CC{\mathbb{C}}
\def\RR{\mathbb{R}}
\def\II{\mathbb{I}}

\begin{document}

\begin{flushright}
    hep-th/9607058 \\ 
    CPTH-S459.0796 \\ 
    July 1996
\end{flushright}
\vspace{.2cm}
\begin{center}
    {\bf Higgs Branch, \Hk quotient and duality \\
    \vspace{.1cm}
     in SUSY N=2 Yang--Mills theories} \\
    \vspace{.3cm}
    I. Antoniadis \footnote{antoniad@orphee.polytechnique.fr}
    and B. Pioline \footnote{pioline@orphee.polytechnique.fr,
                   also at Ecole Normale Sup\'erieure, Paris}\\
    \vspace{.2cm}
    {\it Centre de Physique Th\'eorique
         \footnote{Laboratoire Propre du CNRS UPR A.0014.} \\
         Ecole Polytechnique, F-91128 Palaiseau, France }
\end{center} 
\vspace{.3cm}

\begin{abstract}
Low--energy limits of N=2 supersymmetric field theories in the Higgs branch
are described in terms of a non--linear 4--dimensional \sm on a
\hk target space, classically obtained as a \hk quotient of the
original flat hypermultiplet space by the gauge group. We review in a
pedagogical way this construction, and illustrate it in various examples,
with special attention given to the singularities emerging in the 
low--energy theory. In particular, we thoroughly study the Higgs
branch singularity of Seiberg--Witten $SU(2)$ theory with $N_f$ flavors,
interpreted by Witten as a
small instanton singularity in the moduli space of one instanton on 
$\RR^4$. By explicitly evaluating the metric, we show that
this Higgs branch coincides with the Higgs branch of a $U(1)$ N=2 SUSY theory
with the number of flavors predicted by the singularity
structure of Seiberg--Witten's theory in the Coulomb phase. 
We find another example of Higgs phase duality, 
namely between the Higgs phases of $U(N_c)\; N_f$ flavors
and $U(N_f-N_c)\; N_f$ flavors theories, by using a geometric interpretation
due to Biquard et al. This duality may be relevant for understanding
Seiberg's conjectured duality $N_c \leftrightarrow N_f-N_c$ in N=1 SUSY
$SU(N_c)$ gauge theories.
\end{abstract}

\newpage

\section{Introduction}

In the last couple of years there have been considerable progress towards
an understanding of quantum field theories in the nonperturbative regime,
mainly by focusing on theories with a large amount of symmetries
that can give strong constraints on quantum effects.
Extended supersymmetric theories indeed are prototypes
of quantum field theories where quantum effects are under tight
control, all the more as the number of supersymmetries becomes higher.
A celebrated example is the
N=2 $SU(2)$ with $N_f$ flavors model of Seiberg and Witten
\cite{seiberg/witten:1994.1,seiberg/witten:1994.2}, where special
holomorphy properties and asymptotic behavior in the vector multiplet sector
are powerful enough to allow for the determination
of the full low energy theory on the Coulomb branch. On the other
hand, it is known that the hypermultiplet \ma corresponding to the Higgs
branch receives no radiative corrections \cite{Barbieri:1982,
seiberg/witten:1994.2}. Indeed,
one may assimilate the dynamical scale $\Lambda$ to a vector multiplet,
and use the decoupling of vector multiplets from 
neutral hypermultiplets to argue that this should hold at nonperturbative
level for any $N=2$ super Yang--Mills theory \cite{argyres:1996}.

By contrast, in models derived as low--energy limits
of superstring theories, one in general expects gravitational ($\alpha'$) 
corrections or even perturbative or
nonperturbative quantum effects when the dilaton which determines
the string coupling is part of a hypermultiplet.
In these cases a study of the hypermultiplet sector would
allow new tests of string dualities, which so far rely
mainly on the vector multiplet Coulomb branch structure.
In particular, an understanding and a classification of singularities 
in the rigid limit, possibly in terms of the quantum numbers of the massless 
particles arising at the singular points would give new insights on
string dynamics.

In this work, we investigate how non--trivial hypermultiplet 
manifolds can emerge by reduction of renormalizable N=2 super
Yang--Mills theories to their  
low--energy limit, and how restoration of gauge symmetries and appearance
of massless particles at certain points manifest themselves as singularities
on these manifolds. 
In Section 2 we shall review in some
detail the geometrical {\it \hk quotient} construction of the low--energy
theory, and apply it in Section 3 to models yielding a
(quaternionic) one--dimensional moduli space with orbifold singularities. 
Section 4 will be concerned with the more interesting case of Seiberg--Witten 
$SU(2)$ model with $N_f$ flavors. Its moduli space can be interpreted as the 
moduli space of
$SO(2N_f)$ 1--instantons on $\RR^4$, and exhibits an isolated singularity
at the point where the instanton shrinks to zero size, that we shall
study in detail. We shall also be able to further check the \SW conjectured
singularity structure on the Coulomb phase, by explicitly proving
the equivalence between the Higgs branches of the $SU(2)$ and $U(1)$ theories
for suitable matter contents. 
Finally, we shall find in Section 5 another example of this kind of
Higgs phase duality in the case of $U(N_c)$ theories with $N_f$ flavors.
Thanks to a geometric interpretation of the Higgs branch in terms
of the cotangent bundle of the complex Grassmannian $G_{N_c,N_f}$
due to ref.\cite{biquard:1995}, we shall be able to prove the invariance
of the Higgs branch under \sd. 
This may be relevant for understanding
Seiberg's conjectured duality $N_c \leftrightarrow N_f-N_c$ in N=1 SUSY
$SU(N_c)$ gauge theories, although at first sight this duality does
not seem to generalize to more general gauge groups.

\section{\Hk manifolds and \hk \\ quotients: a reminder}
As is well known, rigid N=2 SUSY theories are constructed out of two types
of multiplets: the vector multiplet comprises  
(together with two Weyl fermions) a gauge field and a complex
scalar that takes its value in a special \kh manifold,
while the hypermultiplet comprises 
two complex scalars taking their values in a \hk \ma
\cite{alvarez-gaume:1981}
(together with two Weyl fermions). In this paper, we shall mainly
be concerned with the hypermultiplet sector.
The following subsections recall the basic facts about \hk
manifolds, triholomorphic isometries and quotient constructions,
hopefully complementing in a pedagogical way the introductions
already existing in the literature
\cite{hitchin:1987,galicki:1987,hitchin:1993,billo:1994,fre:1995}.

\subsection{\Hk manifolds and hypermultiplets} 
A \hk manifold is a n--dimensional riemannian manifold $(M,g)$
with three covariantly constant complex structures $I^x,\,x=1,2,3$,
verifying the quaternion algebra.
We therefore have the following defining properties:
$$\begin{array}{rclr} 
    I^x I^x & =& - 1         & \mbox{(almost complex structure)} \\
    g(I^x X, I^x Y)&=&g(X,Y) & \mbox{(hermiticity)} \\
    {\cal N}^x (X,Y)=0 &\leftrightarrow& 
     \nabla I^x=0              & \mbox{(integrability)}   \\
    I^x I^y&=&-\delta^{xy}1-\epsilon^{xyz}I^z 
                             & \mbox{(quaternion algebra)}
\end{array}$$
where $X,Y$ are vector fields and $N^x(X,Y):=
[I^x X, I^x Y]-[X,Y]-I^x[X,I^x Y]\\-I^x[I^x X,Y]$ 
is the Nijenhuis integrability tensor
\footnote{We use the following standard notations
\cite{kobayashi:1963,nakahara:1990}: $X,Y,\dots$
are vector fields, $[X,Y]$ their Lie bracket, $\Lie{X}$ the Lie derivative
along the vector $X$, $\nabla_X$ the Levi--Civita covariant derivative
along $X$,$X.\phi=\langle d\phi,X\rangle$ the derivative of the scalar 
function $\phi$ along X, $\nabla\phi$ the gradient of $\phi$,
$\langle \omega , X \rangle$ the contraction of
the 1--form $\omega$ with the vector $X$,
$d$ the exterior derivative on forms,
$i_{X}$ the contraction operator with the vector $X$, and $TM$ the tangent
bundle to the \ma $M$.}.

The three hermitian endomorphisms $J^x:=I^x/{2i}$
generate a unitary n--dimensional representation of $SU(2)_\HH$
on the tangent space, which, since $J^x J^x=3/4$, splits into irreducible 
components of spin 1/2 (real dimension 4).
As a consequence, the real dimension $n$ of $M$ has to be a multiple
of four: $N_H=4n$ 
( from now one, we shall by default refer to quaternionic dimension).
Another way to see it is to note that for any vector $X$, the four vectors
$X,I^1 X, I^2 X, I^3 X$ are orthogonal.

{}From these three complex structures one can construct three
non degenerate antisymmetric 2--forms (the \kh forms associated to
the complex structure $I^x$)
\BEQ \omega^{x}(X,Y):=g(I^x X,Y) \EEQ
and since $d\omega^x(X,Y,Z)= {\cal A} \left( g(\nabla_X I^x Y, Z)\right)$,
where ${\cal A}$ is the antisymmetrization operator, one sees that the 
2--forms $\omega^x$ are closed: $M$ is three
times a symplectic manifold. Moreover, by privileging the third direction
in $SU(2)_\HH$, 
it can be checked that $\omega^h:=\omega^2+i\omega^3$ is a holomorphic 
closed form with
respect to $I^1$, so that $(M,I^1,\omega^h)$ is 
actually a holomorphic--symplectic manifold. 

\Hk manifolds can also be characterized by their riemannian holonomy group:
the parallel transport preserves the symplectic form, so that 
the holonomy group
must be contained in $Sp(N_H)\subset SO(4N_H)$. 
In particular, a \hk \ma is Ricci--flat. 

Practically, a way to prove that a \ma is \hk is to exhibit three
closed forms $\omega^x$ and a $SU(2)_\HH$ action on the tangent space
which preserves the metric and such that $\omega^x,\;x=1,2,3$
transforms as a triplet.

\subsection{Triholomorphic isometries and moment map}
The coupling of hypermultiplets to vector multiplets is obtained
by gauging a compact Lie group $G$ of triholomorphic isometries 
of the \hk \ma \cite{andrianopoli:1996}. 
Its Lie algebra $\cal G$ is generated by triholomorphic
Killing vectors, ie vector fields $K$ such that
\BEQ \Lie{K} g = 0, \hspace{1.5cm} \Lie{K}I^{x}=0, \EEQ
the Lie product on ${\cal G}$ being simply the Lie bracket of vector fields.
These two equations imply that $\Lie{K}\omega^{x}=0$.
Since $\Lie{K}= d i_K + i_K d$ and
the forms $\omega^x$ are closed, one obtains
\BEQ d(i_K \omega^{x})=0 \EEQ
This relation can be locally integrated to yield three functions 
$\mm(K)$ on M linearly dependent on the Killing vectors $K$,
ie three {\it moment maps} from M to the dual 
$\cal G^{*}$ of the Lie algebra $\cal G$,
such that
\BEQ i_K \omega^{x} = d \mm( K ) \;\; \forall K \in \cal G
\EEQ
When the second cohomology group of $\cal G$ vanishes
(as occurs for all semi--simple algebras), the constants of
integration can actually be locally
chosen so as to impose the {\it equivariance condition}:
\BEQ \{\mm(K),\mm(L)\}^x = \mm \left( \{ K,L\}^x \right) \EEQ
where $\{\cdot,\cdot\}^x$ is the Poisson bracket
\footnote{The Poisson bracket constructed from a symplectic form $\omega$
associates to two scalar functions $f$ and $g$ on $M$ the function
$\{ f,g\} = \langle df,G \rangle$ where $G$ is the vector dual to 
$dg$ through the form $\omega$, \ie $dg=\omega(G,.)$.}
constructed from the symplectic form $\omega^x$
(for details see for example \cite{fre:1995}).
The left-hand side being globally defined, the moment maps on the right
are then also globally defined. However, 
if $\cal G$ has a nontrivial center, it may happen that the equivariance
condition cannot be imposed, or if it can, some integration 
constants may still remain undetermined. Those will be interpreted
in a SUSY setting as \FI terms.

As it is appears, the moment map is a very general construction on
\sy manifolds with an action preserving the \sy form.
In classical mechanics, it corresponds to the linear momentum
for the case of translations, or angular momentum for the case of
rotations.

\subsection{N=2 SUSY theory, SUSY vacuum and moment map}
Having recalled in some detail the basics of the \hk geometry,
we now come to its implementation in N=2 SUSY field theory.
The general construction was worked out in a geometrical formalism
in ref.\cite{andrianopoli:1996}, and we shall only focus on 
the elements relevant for our study of hypermultiplet moduli space.

The construction starts from a scalar manifold $M=M_V \otimes M_H$
which is the product of a special \kh manifold 
$(M_V,g_{ij^*})$ describing
the vector multiplets and a \hk \ma $(M_H,g_{uv})$ describing the hypers.
The geometry of $M_V$ is defined by a holomorphic section 
$(Y^I,F_I)$ of a $Sp(N_V)$ bundle over $M_V$, through
$g_{ij^*}=\partial_i Y^I \partial_{j^*} {\bar F}_I 
          - \partial_i F_I \partial_{j^*} {\bar Y}^I$ 
The gauge group of dimension $n_V$ acts on the scalars of both sides by 
(tri)holomorphic isometries generated by the Killing vectors
$K^u_I$ and $k^i_I,k^{i*}_I$, and on 
the gauge vectors through its embedding in the
symplectic group $Sp(N_V)$ of electromagnetic duality.
The N=2 SUSY field theory is then defined by a supersymmetric gauged
\sm on $M$, with in particular the scalar potential
\BEQ {\cal V} = e^2 \left( g_{ij^*} k^i_I k^{j^*}_J
                    + 4 h_{uv} K^u_I K^v_J \right) Y^I {\bar Y}^J
              + g^{ij^*} f^I_i {\bar f}^J_{j^*} 
                \sum_{x=1}^{3} {\cal P}_I^x {\cal P}_J^x 
              \; \ge 0 \EEQ 
where $f_i^I=\partial_i Y^I$ is usually invertible
\footnote{In that case, one can choose the coordinates $(z^i,{\bar z}^{i*})$ 
on $M_V$ so that $Y_I=z^i$. However, this is not always possible.}.
This potential usually gives mass to most of the particles, however it may
happen that some directions on $M$ remain unlifted, corresponding classically 
to massless particles. Their dynamics is then given by a non--linear \sm
whose target space is the set $\cal M$ of classical vacua of the theory, 
that is the moduli space of the theory. We shall
restrict our attention to the $N=2$ vacua, given by the equations
\BEQA        k^i_I {\bar Y}^I = 0, \;\;  k^{i*}_I Y^I = 0 \\
             {\cal P}^x_I = 0      \\        
             K^u_I Y^I = 0 \EEQA
These equations can be obtained by requiring ${\cal V}=0$, or equivalently
by demanding that the N=2 SUSY variations of the fermions (in a trivial gauge
background) vanish. 
Several cases may occur:
\begin{itemize}
\item{} 
There may be no solutions at all, in which case the theory 
(classically) breaks N=2 supersymmetry spontaneously. One should then look
for nonzero minima of ${\cal V}$. We shall not pursue this line here,
except to note that the vacua may preserve some (N=1) supersymmetry 
\cite{antoniadis:1996.1,antoniadis:1996.2}.
\item{} 
There may be isolated solutions, so that no massless particles
remain in the low--energy theory.
\item{} 
There may be branches where the fields on the
\hk \ma are fixed at some point while the vectors are free to take 
their values
in a submanifold $\cal M$ of $M_V$. 
This is usually called a {\it Coulomb branch.}
One expects from N=2 SUSY that the low--energy scalar manifold ${\cal M}$, 
given by $k^i_I {\bar Y}^I = k^{i*}_I Y^I = 0$ modulo $G$ is
a special \kh manifold. Although this can be checked in trivial renormalizable
cases, it does not seem to follow directly from these equations in the general 
geometric setting of Ref.\cite{andrianopoli:1996} (it is even not
clear why $\cal M$ should be a complex manifold).
\item{} 
On the contrary, the fields on $M_V$ may be fixed while
the fields on $M_H$ are free to take value in a submanifold.
One usually refers to this branch as a {\it Higgs branch}.
The low--energy manifold ${\cal M}$
is given by $K^u_I Y^I = 0$ and $\mm_I=0$ modulo $G$.
It is easy to see that the first condition preserves the three complex
structures, since they restrict to the tangent space of the
submanifold $K^u_I Y^I=0$0. That the second condition also yields a
\hk manifold is the substance of the \hk quotient construction
\cite{hitchin:1987}
on which we shall dwell in the next
sections: {\it the zero level set of three moment maps modulo the gauge group
$G$ is still \hk}. Note that when $\cal G$ has a non trivial center, 
some constants in the
definition of the moment maps have remained undetermined under 
the equivariance conditions. This freedom in defining the zero level set
modulo elements of the center corresponds to the well--known 
\FI couplings allowed by N=2 SUSY. Thus, the moduli space still
depends on these free parameters.
\item{} 
Finally the fields may take their value in a submanifold
of $M_V \times M_H$, in which case one speaks of a {\it mixed branch}.
It is not clear whether the scalar manifold should still be a product
of a special \kh manifold with a \hk manifold, especially if the gauge
group $G$ acts on $M_V$ and $M_H$ simultaneously, so that the quotient
by $G$ {\it a priori} couples the vectors with the hypers.
\end{itemize}

In the following, we shall focus on the Higgs branch cases, where
the vector multiplet is frozen to zero. We shall make heavy use of
the \hk quotient construction, which we shall now review in detail.

\subsection{Diverse Quotients Constructions}

\paragraph{Symplectic quotient --}
This construction is of daily use in classical mechanics: it is what allows
to fix the total linear momentum (and forget about the center of mass position)
in order to study an isolated system of interacting particles, or what
allows to fix the 
angular momentum (and forget about the actual angular position) in order to
study the motion of a particle in a central force field. The point
here is that by restricting the phase space $M$ ( \ie a \sy manifold )
to the zero level set $M_0$ of the moment map, and taking the quotient
by the symmetry group $\cal G$,  one obtains a manifold 
${\cal M} = M_0 /G$
of real dimension $\dim M - 2 \dim G$ which is again a symplectic manifold
\cite{marsden:1982}. The \sy form $\omega'$ on ${\cal M}$
is the unique \sy form whose 
pull--back on $M_0$ coincides with the restriction
of the original form $\omega$ on $M$ to $M_0$. $\omega'(X',Y')$
is simply defined by taking any two vectors $X,Y$ on $M_0$ that 
project to $X',Y'$,
and letting $\omega'(X',Y')$ :=  $\omega(X,Y)$.
That this does not depend on the particular lift follows from
$\omega(X,K)=\langle d{\cal P}_K ,X \rangle =0$ for $X\in TM_0$.

Note that the construction carries over if one replaces the zero--level
set $M_0$ by the preimage ${\cal P}^{-1}(k)$ under ${\cal P}$ 
of some non--zero invariant
element $k$ of ${\cal G}^*$ (which corresponds to the residual freedom
in the definition of the moment map), or even by 
the preimage of the whole orbit
of an element of ${\cal G}^*$ under the action of $G$ (although
this case does not seem to occur in the setting of SUSY theories).

\paragraph{Riemannian quotient --}
If instead of taking a \sy manifold one starts with a Riemannian manifold
$(M,g)$ with a continuous isometry group $G$ (acting freely on $M$), one can
construct a canonical metric $g'$ on the quotient manifold ${\cal M}=M/G$, by
requiring that the projection $\pi:M \rightarrow M/G$ be a Riemannian
submersion. 
The metric $g'(X',Y')$ on the quotient is obtained by horizontally
lifting $X',Y'$ to $X,Y$, \ie choosing two vectors $X,Y$ orthogonal to the 
Killing vectors $K$ and projecting to $X',Y'$ and then letting
$g'(X',Y'):=g(X,Y)$. The projection from $M_0$ to $\cal M$ is then a
Riemannian submersion. Note that the pull--back of the metric $g'$
to $M_0$ is {\it not} the restriction of $g$ to $M_0$, for this pulled--back
symmetric form is degenerate along the action of the group.

This metric is actually the metric found by considering
the classical low--energy limit of a (non SUSY) gauged \sm on $M$, obtained
by integrating out the massive gauge bosons. Those couple to the scalars
through a gauged metric 
\BEQ g_A(X,X) = g(X+e A^I K_I,X+e A^J K_J) \EEQ
At the classical level, gaussian integration of the gauge bosons
can be carried out (without taking their kinetic terms into account)
by simply minimizing $g_A$ with respect to the corresponding gauge fields
$A^I$. As can easily be seen, this effectively projects  
the vector field $X$ on the subspace orthogonal to the Killing
vectors $A_I$:
\BEQ \langle g_A(X,X) \rangle_A= g(X',X') \EEQ
where $X'$ is the projection of $X$ on the horizontal subspace, thus yielding
the same metric as the quotient construction. 

In more usual field theoretical terms, gauge bosons couple to matter through
gauge currents $J^{\mu}_I$, which are nothing but the pull--back to
ambient space of the 1--form $J_K$ on target space defined by
\BEQ    J_K := g( K, .) \EEQ
Gaussian integration of the gauge bosons of mass $m$ yields a term
$J^{\mu}_I J_{\mu I}$ in the effective lagrangian, which combines with
the original metric to give the effective projected metric $g'$.

Note that if the gauge group does not act freely on the manifold,
it may happen that at some points part of the gauge symmetry
gets restored, \ie the Killing vectors become linearly dependent.
Consequently some gauge bosons remain massless, while the local dimension
of the quotient $\cal M$, given by the dimension of the horizontal
subspace of the tangent space of $M_0$ at the given point, increases.
This corresponds to extra scalars becoming
massless, and to a singularity in the differentiable structure of the
quotient at the corresponding point. It is similar to the singularity
that occurs at the apex of a cone, where the tangent space is exceptionally
3--dimensional while 2D elsewhere. 

\paragraph{\kh quotient --}
Since a \kh \ma is a special case of \sy manifold, 
the \sy quotient construction
applies and yields a \sy form $\omega'$ on the quotient $M_0/G$ 
if the action of
$G$ is symplectic (ie. $\Lie{K}\omega=0$). If moreover $K$ is a Killing
vector of $M$ (ie. $\Lie{K} g=0$, so that $\Lie{K}I=0$), it is also a 
Killing vector on $M_0$
with the restricted metric $g$, so that $M_0/G$ receives a metric $g'$ through
the Riemannian quotient construction. One can also check that the complex
structure $I$ restricts to the horizontal subspace of $TM_0$, 
and thus descends to a complex structure $I'$ on the quotient.
The compatibility of $g',I',\omega'$ ensures that
$\cal M$ is indeed a \kh manifold.

That $\cal M$ is a complex manifold can more easily be seen by noting
that it coincides with the quotient of $M$ by the action of the complexified
group $G^{\CC}$ generated by the holomorphic vector fields $K\pm i\;IK$.
Indeed, the orbit of a generic point $x$ of $M$ under the imaginary part 
$e^{i{\cal G}^*}$ of the complexified gauge
group generically intersects the zero--level submanifold in exactly one point
$x_0=g_x x$
(since the flow corresponding to the vectors $IK$ is orthogonal to $M_0$).
On the other hand two points $x,y$ 
in $M$ equivalent under the action of $G^\CC$
are mapped under this procedure into two points $x_0,y_0$ on $M_0$
equivalent under the action of $G$ itself. The holomorphic symplectic
structure on ${\cal M}=M_0/G$ is then the same as the quotient holomorphic
symplectic structure on $M/G^\CC$. 
This is summarized in the following diagram:
$$\begin{array}{ccc}
  M          & \stackrel{g_x}{\longrightarrow}            & M_0 \\
  \downarrow &                                           & \downarrow \\
  M/G^\CC	     & \stackrel{Hol.Sympl}{\longleftrightarrow} & M_0/G
  \end{array}
$$
Actually, since $G^\CC$ is non compact, the quotient is in general ill-defined
at some points, and we should restrict $M$ to the set of 
{\it stable points}, \ie those which have a point in $M_0$ in the closure
of their orbit under $G^\CC$ (This is pedagogically explained
in ref.\cite{witten:1993}).

One can use this mapping to pull the \kh metric from the \kh 
quotient back to the complex manifold $M$, and a formula
for the resulting \kh potential has been given in
\cite{biquard:1995}, exploiting an idea of \cite{hitchin:1987}:
\BEQ 
K'(x) = K( g_x x) + {1\over 4\pi} \ln|\chi(g_x)|^2 
\label{biquard}
\EEQ
The second term is only present in presence of \FI terms, \ie when one 
considers, instead of the zero--level set $M_0$ of the moment map, the
preimage of a G-invariant element $k$ of ${\cal G}^*$. This element
can be seen as the differential of a character $\chi:
G \rightarrow U(1)$ at the unity of $G$:
$d\chi=-2\pi i k$. In equation (\ref{biquard}), $\chi$ is naturally extended
to the complexified group $G^\CC \rightarrow \CC^*$.

\paragraph{\Hk quotient --}
This construction can now be generalized to the case of \hk manifolds
\cite{hitchin:1987,galicki:1987}.
Here we define $M_0$ as the intersecting zero level set of the three 
moment maps: $M_0 = \{z\in M / \mm=0, \, x=1,2,3\}$, and ${\cal M} = M_0/G$.
${\cal M}$ is thus of real dimension $\dim M - 4\dim G$, and as it turns
out still \hk. To see why it is a special case of the previous construction,
note that $M_h = \{z\in M/ {\cal P}^1={\cal P}^2=0\}$ is a 
complex submanifold of
$(M,I^1)$, inheriting its \kh structure from $M$, and stable under the
action of $G^\CC$. The previous construction then yields a \kh structure
on its \kh quotient $\cal M$. The same can be done by privileging the
two other complex structures in turn, and the three \kh structures on
$\cal M$, as it turns out, make it into a \hk \ma. The three complex
structures on $\cal M$ are simply the restrictions of the original ones
to the horizontal subspace of the tangent space of $M_0$, as this subspace
is stable under $I^x,\,x=1,2,3$. In the cases where there remains 
unbroken gauge symmetry, the local dimension of the quotient increases 
by a multiple of four (since adding a vector X to the horizontal subspace
automatically brings in $I^x X,\, x=1,2,3$, and these four vectors are
linearly independent as well as independent from the original ones), 
and correspondingly extra hypermultiplets become massless, one for each
unbroken gauge symmetry. This is the N=2 version of the Higgs effect,
the gauge vector multiplet becoming massive by ``eating'' a matter
hypermultiplet. As a consequence, the index $N_V-N_H$ is a constant over
the Higgs branch
\footnote{A vector multiplet can also become massive when a central charge
appears in the SUSY algebra, but no example of this involving hypermultiplets
has been found.}.

As a holomorphic--symplectic manifold, the \hk quotient also coincides
for stable points with the holomorphic quotient of the complex submanifold
$M_h$ by the complexified group $G^\CC$. One may ask whether it can
be obtained directly form $M$ by quotient by an hypothetical ``quaternionized''
group $G^\HH$. The problem here is that there is no good Lie algebra 
structure on the tensor product ${\cal G} \otimes \HH$.

\subsection{Renormalizable N=2 SUSY field theories}
In the following, we shall be interested in low--energy effective theories
corresponding to microscopic renormalizable gauge theories, \ie corresponding
to a non linear \sm with flat target space. The way to obtain these flat
manifolds is to take a $\CC$--vector 
space $V$ of complex dimension $N_H$, acted upon by a 
linear unitary representation of a Lie group $G$ 
(which has to be a subgroup of $Sp(N_H))$.
G acts on its dual $V^*$ by the contragradient representation, and we
choose $M= V \oplus V^*$. Letting ${Q^i, \tilde{Q}_i, \, i=1\dots N_H}$ be
the coordinates on M, we have the \hk structure
\BEQA   \omega^x = \pmatrix{dQ^+_i & d\tilde{Q}_i } \sigma^x \wedge
             \pmatrix{ dQ^i \cr d\tilde{Q}^{i+} }      
	\hspace{1.5cm}
        g =  \pmatrix{dQ^+_i & d\tilde{Q}_i } \otimes
             \pmatrix{ dQ^i \cr d\tilde{Q}^{i+} }      
\label{flatmetric}
\EEQA
where the $\sigma^x$ are the Pauli matrices. The $SU(2)_\HH$ action on
the cotangent bundle is such that $\pmatrix{ dQ^i & d\tilde{Q}^{i+} }$
transforms as a doublet. This is not to be mistaken with an extra $SU(2)_R$
isometry that acts on the flat \hk space itself, under which the coordinates
$\pmatrix{ Q^i & \tilde{Q}^{i+} }$ themselves transform as doublets
\footnote{This $SU(2)_R$ is actually nothing but the action of the 
unit quaternions on the quaternionic vector space $V\oplus V^*$.}
(while the vector multiplet scalars would be singlets). 
Although this isometry is not triholomorphic, it still generates a
symmetry in the full gauged theory, if one asks that the gauginos
transform as doublets while the hyperinos are singlets.

The Killing vectors associated to $G$ read 
\BEQ   K_I=  \pmatrix{\partial_{Q^+_i} & \partial_{\tilde{Q}_i}} T^{i}_{jI}
              \pmatrix{ Q^j \cr \tilde{Q}^{j+} }      
              -
              \pmatrix{Q^+_i & \tilde{Q}_i }  T^{i}_{jI}
              \pmatrix{\partial_{Q^j} \cr \partial_{\tilde{Q}^{j+} }}      
\EEQ
where $T^i_{jI}$ is the antihermitian representation of the generator
associated to $K_I$,
while the equivariant moment maps are
\BEQ   \mm_{K_I} = \pmatrix{Q^+_i & \tilde{Q}_i} \sigma^x  T^{i}_{jI}
                   \pmatrix{ Q^j \cr \tilde{Q}^{j+} }      
\EEQ
up to \FI terms for the generators in the center of $\cal G$.

In N=1 superfield formalism, the moment map ${\cal P}^3$ simply corresponds
to the D--term coming from  the canonical kinetic terms, while the
(anti)holo\-mor\-phic moment maps ${\cal P}^1 \pm i{\cal P}^2$
correspond to the
F--terms of the vector multiplet induced by the superpotential
\BEQ   W = \tilde{Q}_i  T^{i}_{jI} \Phi^I Q^j \EEQ
where $\Phi^I$ stands for the chiral superfield of the vector multiplets.
For short, we shall call D--flatness (resp. F--flatness) the conditions
${\cal P}^3=0$ (resp. ${\cal P}^1 + i{\cal P}^2=0$).

One may ask how the mass terms appear in this formalism: the only way
to introduce them is to consider them as frozen N=2 vector multiplets
gauging a flavor group. In order to be consistently frozen,
their fermionic components should be invariant under N=2 symmetry, ie
one should have $k_I^{i} Y^I=k_I^{i*} Y^I=0$. 
These vector multiplets would then give extra contribution to the 
scalar potential, and would change the vacuum equations 
$K_I^u Y^I=0$. They would also bring in extra moment map equations
which however
may be dropped by choosing an infinite metric on the vector
multiplet manifold in the direction corresponding to the flavor
vector multiplets.

In the following we will present ex\-plicit calcu\-la\-tions of \hk
quotients of some examples of these flat manifolds, and study 
the singularities that emerge. 

\section{1--dimensional examples and ADE singularities}
\Hk quotient manifolds of dimension 1 are the most easy to study, since
it is often possible to explicitly parameterize $M_0$ modulo the gauge
group. Moreover, 1--dimensional asymptotically locally euclidean (ALE)
\hk manifolds
have been under active investigation as gravitational backgrounds
for general relativity, and have been completely classified in
terms of resolutions of the quotient of $\CC^2$ by
a discrete subgroup of $SU(2)_R$ \cite{kronheimer:1989}. 

\subsection{U(1) Gauge theory and Eguchi--Hanson gravitational instantons}
Let us consider a renormalizable theory with $N_f$ 
hypermultiplets $(Q_i,\tilde{Q}_i$ with 
charges $e_i^\alpha$ under $N_c$ U(1) vector multiplets $\Phi^\alpha$. The 
Higgs branch vacuum equations, as read on the superpotential
$W=e^\alpha_i \tilde{Q}_i \Phi^\alpha Q_i$, are
\BEQA   e^\alpha_i \tilde{Q}_i Q_i = \xi^{\alpha} \in \CC  \\
        e^\alpha_i ( Q^+_i Q_i - \tilde{Q}_i \tilde{Q}^+_i ) 
                     = \nu^\alpha \in\RR
\EEQA
where $\xi$ and $\nu$ denote the three \FI parameters.
In the case $N_f=N_c+1$ and $e_i^\alpha$ of maximal rank, one expects 
a dimension 1 Higgs branch with all the $U(1)$'s broken by expectation values
of the hypermultiplets. For vanishing \FI terms, we 
can actually give an explicit parameterization of the manifold of solutions
after some changes of basis. First, modulo interchange of hypermultiplets
we can assume that the square submatrix $e^\alpha_i, \, \alpha,i=1..N_c$
is invertible, and act by $GL(N_c,\RR)$ combinations of $U(1)$'s to turn it
into unity (whereas this does not preserve the scalar potential, it
definitely preserves the flat directions). This leaves the values
$e_{N_f}^\alpha=q^\alpha$. If any of these is zero, 
1 vector and 1 hyper decouple.
Otherwise we can rescale the U(1)'s to achieve 
\BEQ e_i^\alpha= \pmatrix{ 1/q^1&  &  &  &  1 \cr
                            & 1/q^2&  &  &  1 \cr
                            &  &\ddots& & \vdots   \cr
                            &  &   &1/q^{N_c}&  1}
\EEQ
and rescale the hypers to choose $q^\alpha=1$. The Higgs branch is then
para\-metri\-zed by 
\BEQ \pmatrix{Q_i \cr \tilde{Q}_i} =
     \sqrt{q_i} \pmatrix{a \cr b}, \;\; 
     \pmatrix{Q_{N_f} \cr \tilde{Q}_{N_f}} =   
                        \pmatrix{-b \cr a }, \hspace{.5cm}(a,b)\in \CC^2   \EEQ
The gauge current
vanishes for this parameterization, so we actually have a slice of $M_0$
orthogonal to infinitesimal gauge transformations.
However, there may still remain a discrete subgroup $\Gamma$ of the $U(1)$'s
relating some $(a,b)$'s, by which we should quotient $\CC^2$. Note that
the precise subgroup depends crucially on the charges and is not invariant
under linear redefinitions of $U(1)$'s. For the previous $q^\alpha=1$ charge
assignment, a $U(1)^{N_c}$ transformation 
$(e^{i\theta_1},\dots,e^{i\theta_{N_c}})$ on $(Q_i,\tilde{Q}_i)$ can be
reabsorbed in a change of $(a,b)$ for
\BEQ
  e^{i\theta_1} = \dots = e^{i\theta_{N_c}} =
  e^{-i\theta_1 \dots -i\theta_{N_c}} \EEQ
that is $\theta_i=\theta, \; (N_c+1)\theta \equiv 0 \pmod{2\pi}$.
The moduli space is then $\CC^2/\ZZ_{N_f}$, where the discrete group acts as
$(a\, e^{2i\pi/{N_f}},b\, e^{-2i\pi/{N_f}}) \equiv (a,b)$.

On the other hand, for the case of two hypers of charges
$(p,q)$ under one $U(1)$, the gauge transformation acts as
\BEQ \pmatrix{ Q_1 & Q_2 \cr \tilde{Q}_1 & \tilde{Q}_2 }
     = \pmatrix{ a & -b \cr b & a} \longrightarrow
     \pmatrix{ a e^{ip\theta} & -b e^{iq\theta} \cr 
               b e^{-ip\theta}& a e^{-iq\theta}}
\EEQ
so that it can be reabsorbed in a change of $(a,b)$ for 
$(p+q)\theta \equiv 0 \pmod{2\pi}$. The corresponding subgroup is 
$\ZZ_r$ where $r$ is the smallest integer so that $pr/(p+q)$ is integer,
that is $r=\mbox{lcm}(p,p+q)/p$.

As for the metric on $\cal M$, it is anyway obtained by pulling back 
the metric (\ref{flatmetric}) of the unconstrained $Q,\tilde{Q}$ fields:
\BEQ ds^2 = (1+\sum | q_i |) (da da^+ + db db^+) \EEQ
This is still the flat metric on $\CC^2/\Gamma$, but due to the quotient
the space is not flat, but rather has an orbifold singularity in curvature at
the origin (its holonomy group is the discrete group $\Gamma$ rather 
than the trivial group).  

This space (for $q_i=1$) is actually the multi--Eguchi--Hanson
gravitational instanton 
\cite{eguchi/hanson:1978,eguchi/hanson:1979},
in the limit where the $N_f$ instantons
collapse to one point. Switching on \FI terms generically removes the
points with unbroken gauge symmetry and therefore yields a smooth
manifold, indistinguishable from the singular one at long distance.
The orbifold singularity for vanishing \FI terms is resolved into
a family of $N_c$ intersecting two spheres, and one can retrieve the value
of the $3N_f-3$ real \FI parameters by integrating the three closed 
\hk forms on these spheres \cite{hitchin:1994}
\footnote{In Ref.\cite{duistermatt:1982} it was proved that the 
\kh classes of the \hk forms
are actually linear in the \FI terms, so that the periods of these forms
yield the \FI terms for suitable normalizations.} 
On the other hand, there are also $3N_f-3$ real parameters in the
multi--Eguchi--Hanson specifying the relative positions of the instantons,
and one can check that those are exactly the \FI parameters.
Setting $n$ triplets of \FI terms to zero is then equivalent to shrink $n$
2--spheres to zero, or to make $n$ instantons collapse at the same point.
The space then looks locally like $\CC/\ZZ_n$.

\subsection{$SU(N_c)$ with $N_f=N_c$ flavors gauge theory}
\label{sunc}
Another easily workable example is the case of N=2 SQCD $SU(N_c)$
with $N_f=N_c$ flavors, where we also expect
a 1--dimensional Higgs phase, among other phases with partially
restored symmetry. The Lie algebra of $SU(N_c)$ has a trivial center,
so no \FI terms are available, and
the F-- and D--flatness equations take the form
\BEQA  \tilde{Q} Q \propto \II_{N_c} \\
       Q^+ Q - \tilde{Q} \tilde{Q}^+ \propto \II_{N_c} \EEQA
where in the general case $(N_f,N_c)$ case $Q$ (resp $\tilde{Q}$) is a 
$N_c\times N_f$ matrix (resp.  $N_f\times N_c$) (For this subsection
we restrict to $N_f=N_c$). 
The right-hand side can be seen as the contribution
of the Lagrange multiplier imposing that the vector multiplet is traceless.

An obvious and gauge--orthogonal solution is $Q=a\,\II_{N_c}, 
\tilde{Q}=b\,\II_{N_c},\-
\, (a,b)\in \CC^2$, but here again $(a,b)$ and 
$(a\, e^{2i\pi/N_c},b\, e^{-2i\pi/N_c})$ are gauge equivalent,
so the moduli space is really ${\cal M}=\CC^2/\ZZ_{N_c}$, again with a
flat metric and an orbifold singularity at the origin, just like
the $U(1)^{N_c}$ with $N_c+1$ flavors case. This example
shows that there is no hope to identify the restored gauge symmetry
at a singularity by inspection of a Higgs branch only.
However, for this model we have only looked at a part of the moduli space,
namely the baryonic branch in the terms of 
ref.\cite{argyres:1996} (so called because
the baryonic operators $\det Q$,$\det\tilde{Q}$ are non vanishing), and
there are also a variety of non baryonic branches meeting the latter
at the origin. The complete structure of those branches may be sufficient
to characterize the singularity at the origin.

\subsection{Kronheimer--Nakajima construction}
So far we only have seen $\ZZ_n$ type of orbifold singularities emerge.
As already mentioned, ALE 1--dimensional \hk manifolds are 
completely classified, and are desingularizations of quotients
of $\CC^2=\HH$ by a discrete subgroup $\Gamma$ of $SU(2)_R$,
the latter being in one--to--one correspondence with simply laced ADE 
Dynkin diagrams.
It should therefore be possible to obtain any kind of ADE orbifold
singularity by looking at low--energy limits of suitable N=2 SUSY field
theories.

This construction has actually be found by Kronheimer
\cite{kronheimer:1989},reviewed for physicists
in \cite{anselmi:1994,billo:1994}, and
then extended by Nakajima \cite{nakajima:1994} in the formalism 
of {\it quiver varieties}, more natural for a SUSY field
theory interpretation. This formalism has recently found its way
in field theories through the study of D-branes
\cite{douglas:1996}, so we shall briefly
review the construction.

For any oriented diagram, \ie a collection of points and arrows joining
(some of) them, we associate to each point a gauge group $U(N_i)$ with
the corresponding vector multiplets, and to each arrow $i\rightarrow j$
hypermultiplets in the representation $({\bf N_i},{\bf\bar N_j}$ of 
$U(N_i)\times U(N_j)$
\footnote{This field content is typical of the states found from open strings
in the background of D-branes, which carry Chan--Paton gauge 
indices labeling the branes on which each end of the string lives.}
. This defines a N=2 field theory with 
gauge group $\prod U(N_i)$ (one of the U(1) being decoupled)
and a Higgs branch which is a \hk manifold of dimension
computable in terms of the $N_i$'s and the connection matrix
of the diagram (non singular for generic values of the \FI terms).
In particular, 1--dimensional manifolds occur when one chooses
a simply--laced extended Dynkin diagram and associates to each point
the Coxeter number of the corresponding representation
(the corresponding diagrams with the Coxeter numbers can be
found in ref.\cite{billo:1994},fig.1 and 2). The resulting 
manifold is then a ALE gravitational instanton,
asymptotically equivalent to $\CC^2/\Gamma$.

For instance, a $D_4$ singularity would be obtained by studying the
Higgs branch of a $SU(2)\times U(1)^4$ theory with 4 doublets of hypers
each one of charge one under a different $U(1)$. Unfortunately, we
could not find any parameterization of the Higgs branch that would
exhibit the $D_4$ discrete action. A $E_8$ singularity could be obtained
at the price of a rather heavy gauge group: 
$$ U(1)\times (U(2)\times U(3))^2\times U(4)\times U(5) \times U(6)$$
 
\section{Seiberg--Witten theory and small instanton singularity}
Models with $N_H\ge 2$ are usually rather difficult to solve unless
they possess some special flavor symmetries. The case of N=2 SUSY
$SU(2)_G$ gauge theory 
\footnote{We use a subscript G to distinguish the $SU(2)$ gauge group
from the other $SU(2)$'s that will occur in the following. In the context
of hypermultiplets it is more useful to think of this $SU(2)_G$ as a
$Sp(1)_G$.}
with $N_f$ flavors is particularly favorable, since the 
pseudoreality of the $SU(2)_G$ representation ${\bf 2}$ implies that
the symmetry group is enhanced from $SU(N_f)\times SU(2)_G\times SU(2)_R$ to 
$SO(2N_f)\times SU(2)_G \times SU(2)_R$. The Higgs branch is then 
specified by only one real parameter up to symmetries. Moreover,
a nonzero expectation value for a single doublet of $SU(2)$
\footnote{Although for $N_f=1$ there is no Higgs branch, as we shall
see in the following.} already breaks 
the gauge symmetry completely, so gauge symmetry
enhancement can only happen at the origin in field space, and a single
isolated singularity is expected. The Higgs branch of this model has 
actually been briefly
worked out in \cite{seiberg/witten:1994.2}, 
and subsequently in more detail in the context of
small instantons in heterotic $SO(32)$ string theory
\cite{witten:1996}. Here we shall 
concentrate on the study of the singularity occurring at the origin,
after recalling Seiberg and Witten's description of the moduli space.

\subsection{$SO(2N_f)\times SO(4)$ symmetry and solutions of the vacuum
equations}
Using the same notations as in subsection \ref{sunc}
and embedding $SU(N_f)\subset SO(2N_f)$
through $A+iB \hookrightarrow \pmatrix{ A & B \cr -B & A}$, 
the hypermultiplets can be recast in a pseudoreal 
$({\bf 2N_f},{\bf 2}, {\bf 2})$ representation $q_{I\alpha}^a$ of 
$SO(2N_f)\times SU(2)_G \times SU(2)_R$:
\BEQ  q_{I1}^a=\pmatrix{Q_i^a + \epsilon^{ab}\tilde{Q}^i_b \cr
                        iQ_i^a -i \epsilon^{ab}\tilde{Q}^i_b } 
   \hspace{1.5cm}
   q_{I2}^a=\pmatrix{\epsilon^{ab}Q^{i+}_b - \tilde{Q}^{a+}_i \cr
                     -i\epsilon^{ab}Q^{i+}_b - i\tilde{Q}^{a+}_i }
\EEQ    
where the indices $I,\alpha,a$ label the defining representations of 
$SO(2N_f)$,\- $SU(2)_R$,\-$SU(2)_G$. The pseudoreality condition reads
\footnote{
Should we have taken matter in the adjoint representation
of $SU(2)_G$, there would have been no such reality
condition to be imposed on the $SO(2N_f)\times SU(2)_G \times SU(2)_R$
$({\bf 2N_f},{\bf 3}, {\bf 2})$ representation to cut the number of degrees
of freedom by half, so that a $SO(2N_f)$ enhancement could not occur.
On the other hand, it would still take place if one chooses a gauge
group $Sp(N_c)_G$ rather than $SU(2)_G=Sp(1)_G$ with $N_f$ hypers
in the fundamental representation; indeed,
as noted in Ref.\cite{witten:1996}, one could use the antisymmetric
tensors of $Sp(N_c)$ and $SU(2)_R$ to impose a reality condition
on the $({\bf 2N_f},{\bf N_c}, {\bf 2})$ representation of
$SO(2N_f)\times Sp(N_c)_G \times SU(2)_R$.
}:
\BEQ (q_{I\alpha}^a)^+ = \epsilon^{\alpha\beta} \epsilon_{ab} \;
                         q_{I\beta}^b \EEQ
In this formalism, the metric and moment maps translate into
\BEQ ds^2 = \epsilon^{\alpha\beta} \epsilon_{ab} \;
            dq_{I\alpha}^a \otimes dq_{I\beta}^b  \hspace{1.5cm}
     {\cal P}^{x(ab)} = q_{I\alpha}^{(a} (\sigma^x)^\alpha_{\;\beta}
                      \epsilon^{\beta\beta'} q_{I\beta'}^{b)}   \EEQ
where the adjoint representation of $SU(2)_G$ is written as a 
symmetric tensor ${\cal P}^{(ab)}$. Vanishing of the three moment maps
requires
\BEQ q_{I\alpha}^a \; q_{I\beta}^b \propto 
                         \epsilon_{\alpha\beta} \epsilon^{ab} \EEQ
This can be more easily exploited if one uses the decomposition
$SO(4)=SU(2)_G \otimes SU(2)_R/\ZZ_2$ 
\footnote{This generalizes to the $Sp(N_c)$ case by using the embedding
$Sp(N_c)_G\times SU(2)_R \subset SO(2N_c)$ under which ${\bf 2N_c}$
decomposes precisely as $({\bf N_c},{\bf 2})$. The flatness conditions
$q_{I\alpha}^a q_{I\beta}^b \propto \epsilon_{\alpha\beta}$ would not
however have a simple interpretation in terms of $SO(2N_c)$.}
under which the pseudoreal $({\bf 2},{\bf 2})$
representation corresponds to a real vector of SO(4) through
\BEQ q_{I}^{\mu} := (\sigma^\mu)_{\alpha\dot{\alpha}} 
                    \epsilon^{\dot{\alpha}\dot{\beta}} 
                    q_{I\dot{\beta}}^{\alpha}
\EEQ
where $\sigma^\mu$ are the generalized 4 Pauli matrices
(see for instance Ref.\cite{wess/bagger:1992}).
In this formalism the flatness conditions and metric read
\BEQA  q_I^\mu q_I^\nu \propto \delta^{\mu\nu} \\
       ds^2= dq_I^\mu \; dq_I^\mu
\label{metricso}
\EEQA
A point on the flat directions is thus given by 4 orthogonal real vectors of
$\RR^{2N_f}$, each of the same undetermined length $\rho$. 
$SO(2N_f)$ acts irreducibly
\footnote{This is not quite true for $2N_f=4$ where the orientation of
4 vectors in $\RR^4$ distinguishes two connected components 
(the so called baryonic and nonbaryonic branches) that can be mapped
to each other by a $O(2N_f)$ ($Q_1^a \leftrightarrow -\epsilon^{ab}
\tilde{Q}^1_b$) or $O(4)$ parity transformation. The action of $SO(2N_f)$
is then irreducible on each branch.}
on the bases of 4 vectors of a given length, so
we can bring the four vectors along the first four directions of $\RR^{2N_f}$
and recover the general solution by a $SO(2N_f)$ rotation. We can
consequently parameterize the Higgs branch as
\BEQ  q_I^\mu = \Omega_{2N_f} \pmatrix{ \rho &     &    & \cr
                                      &\rho &    & \cr
                                      &     &\rho& \cr
                                      &     &    & \rho \cr \cr \cr},
      \;
 \begin{array}{rcl}
      \Omega_{2N_f} &\in& SO(2N_f) \\
      \rho          &\in& \RR^+          \end{array}
\EEQ
However, a subgroup $SO(2N_f-4)$ of $SO(2N_f)$ lets the solution unaffected,
and furthermore one should identify configurations differing by the action of 
the gauge group $SU(2)_G \subset SO(4)$. 
One therefore obtains as a moduli space
\BEQ {\cal M} = \RR^+ \times {SO(2N_f) \over SO(2N_f-4) \times SU(2)_G} \EEQ
where $SO(2N_f-4) \times SU(2)_G$ acts in $SO(2N_f)$ as
\BEQ \Omega_{2N_f} \stackrel{\equiv} \longrightarrow \Omega_{2N_f} .
                          \pmatrix{\II_4& \cr & \Omega_{2N_f-4}} .
			  \pmatrix{\Omega_4 & \cr & \II_{2N_f-4}}, 
\EEQ
with $\Omega_{2N_f-4}\in SO(2N_f-4)$ and $\Omega_4\in SO(4)$ the embedding of
$SU(2)_G$ in SO(4).

To evaluate the metric on this space
\footnote{Note that there is not a unique $SO(2N_f)$ invariant metric
on the coset $SO(2N_f)/\-SO(2N_f-4)\times SU(2)_G$}, we first
pull the flat metric (\ref{metricso}) back on
$\RR^+ \times SO(2N_f)$:
\BEQA ds^2 = tr\left( d(\rho\Omega_{2N_f})
                 \pmatrix{\II_4 \cr \cr } \pmatrix{\II_4 & }
                 d(\rho\Omega_{2N_f})^{t} \right)  \\
           = d\rho^2 + \rho^2 \tr \left(\Omega_{2N_f}^t d\Omega_{2N_f}
                 \pmatrix{\II_4 & \cr & } d\Omega_{2N_f}^t 
                 \Omega_{2N_f} \right)
\EEQA
Here we note that the scale fluctuations $d\rho$ are orthogonal to the 
fluctuations of $\Omega_{2N_f}$, and that the fluctuations along
$SO(2N_f-4)$ are effectively projected out. Then we should retain only
the fluctuations of $q$ orthogonal to the gauge group $SU(2)_G$.
Scale fluctuations are already orthogonal to $SU(2)_G$, so one simply
has to project the fluctuations $d\Omega_{2N_f}$ on the subspace
orthogonal to the subalgebra $SU(2)_G$ for the scalar product
$\tr \left( d\Omega_{2N_f} \pmatrix{\II_4 & \cr & } 
     d\Omega_{2N_f}^t \right) $:
\BEQ
   ds'^2 = d\rho^2 + \rho^2 \tr \left(\Omega_{2N_f}^t d\Omega_{2N_f}
                               \right)_\perp
                 \pmatrix{\II_4 & \cr & } 
                    \left( d\Omega_{2N_f}^t \Omega_{2N_f}
                 \right)_{\perp}
\label{metricsu2}
\EEQ

This a special case of a {\it warped product} $M\otimes_{f^2} M'$
of two Riemannian manifolds $(M,ds^2)$ and $(M',ds'^2)$, 
that is a Riemannian
manifold $M\times M'$ with metric $d\tilde{s}^2= ds^2 + f^2(x) ds'^2$
where $f(x)$ is a function of the coordinates on $M$ only.
These manifolds have actually been under investigation in the context 
of Einstein spaces  \cite{besse:1986}
and have yielded numerous examples of non--homogeneous
Einstein manifolds, though non compact. The case where $\dim M=1$ has
in particularly been completely worked out, and it is 
known that Ricci--flat warped products are obtained only for $f(x)=x$
and $M'$ an Einstein manifold of definite Einstein constant. 
This is precisely the case here.

\subsection{Conical singularity}

Singularities on such manifolds may arise when $f(x)$ vanishes, and this
actually occurs in the case at hand when $\rho$ vanishes. It was noticed
by Witten \cite{witten:1996} that the Higgs branch 
we are considering actually describes
the moduli space of one $SO(2N_f)$ instanton in $\RR^4$, as one
learns from the general ADHM construction
\cite{atiyah:1978}. The four
directions $q_I^\mu$ describe the embedding of the $SU(2)$ describing
the instanton in $SO(2N_f)$, while $\rho$ specifies its size. The
singularity at the origin thus corresponds to the zero size limit
of the instanton, and signals the appearance of a ``nonperturbative''
$SU(2)$ gauge symmetry enhancement. 

The warped product structure makes it fairly easy to study the 
singularity that occurs at the origin, since it does not come from either
of the components but only of the function $f(x)$ that couples them.
The Riemann tensor of a warped product can be easily expressed in
terms of the Riemann tensors of $M$ and $M'$ (expressions can be found in
\cite{besse:1986}, but we give in appendix 
another version of them with more conventional
notations), and it is found that the only non vanishing component is
obtained when all the vectors are on the homogeneous side:
\BEQ 
  \tilde{R}(X',Y')Z'=R'(X',Y')Z'+ \left( g'(X',Z')Y'-g'(Y',Z')X' \right)
\EEQ
that is, the only effect of the warping is to add a constant 
negative curvature term to the Riemann tensor on the homogeneous side.
As a consequence, for closed paths at fixed value $\rho$
arbitrarily close to zero, the holonomy can be calculated in terms of
$M'$ only, using the previous expression as an effective Riemann tensor,
and is independent of $\rho$. Except in the case where special cancellation
between the two terms occurs, the holonomy remains non trivial when the path
shrinks to zero, implying a singularity in the Riemann tensor at this
point, even though the components of the Riemann tensor do not
show any divergence. Cancellation can only occur when $M'$ is
of constant positive curvature, \ie locally a sphere. This is actually
what happens for $2N_f=4$, since 
\BEQ {SO(4)\over SU(2)_G} \equiv {SU(2)_R \over \ZZ_2} 
                          \equiv {S_3 /\ZZ_2}             \EEQ
so that the holonomy group around the origin is a discrete group $\ZZ_2$,
corresponding to an orbifold singularity. This agrees with the
result ${\cal M}=\CC^2/\ZZ_2$ of the previous section. Note that in general
the singularity is much worse than an orbifold singularity, since
the local holonomy group is not even discrete.

\subsection{Global symmetries on the Higgs branch}
{}From the previous formulation of the Higgs branch, it is easy to determine
the global symmetry breaking pattern at a given point in the moduli space
(as was already done in the
original paper by Seiberg and Witten). The four vectors of $SO(2N_f)$ break
the flavor symmetry down to $SO(2N_f-4)$, however $SO(4)$ rotations in 
the space of the four vectors can actually be compensated by orthogonal
linear combinations of the same vectors (this is only true because they are 
orthonormal), \ie by $SU(2)_G\times SU(2)_R$ rotations. 
The remaining global symmetry group is thus
\BEQ  SO(2N_f-4) \times SU(2)_{G'} \times SU(2)_{R'} \EEQ
where $SU(2)_{G'}$ is the diagonal group of $SU(2)_G$ times a $SU(2)$
subgroup of $SO(4)\subset SO(2N_f)$, while $SU(2)_{R'}$ is the diagonal group 
of $SU(2)_R$ times the other $SU(2)$ subgroup of $SO(4)$.
This symmetry group acts on the tangent space of ${\cal M}$, which splits
into real irreducible representations
\BEQ ({\bf 2N_f-4},{\bf 2},{\bf 2}) \oplus 
     ({\bf 1},{\bf 1},{\bf 3}) \oplus
     ({\bf 1},{\bf 1},{\bf 1})  \EEQ
as obtained by decomposing the adjoint representation of $SO(2N_f)$
under the unbroken group and forgetting the adjoint representations
of $SO(2N_f-4)\times SU(2)_G$ by which we quotient. This
corresponds to the representations of the massless particles for the
given point in moduli space. The representations of the fermions can also
be worked out, taking into account their different quantum numbers under
$SU(2)_R$, yielding
\BEQ  ({\bf 2N_f-4},{\bf 2},{\bf 2}) \oplus 
     ({\bf 1},{\bf 1},{\bf 2}) \oplus
     ({\bf 1},{\bf 1},{\bf 2})  \EEQ
The unusual quantum numbers of the massless particles under $SU(2)_{R'}$
should cause no surprise: the action of $SU(2)_{R'}$ is distinct from
that of the $SU(2)_\HH$ generated by the three complex structures, which is
an isometry of the tangent space but not a symmetry of the theory in general.

\subsection{N=2 duality and Higgs branches}
As is now well known, the $U(1)$ Coulomb phase of the \SW model 
$SU(2)_G$ with $N_f$ flavors presents 
singularities where hypermultiplets charged
under $U(1)$ become massless. From these points Higgs branches emerge
corresponding to giving vacuum expectation values to this multiplet,
and these branches should actually be the same as the Higgs branches
of the microscopic $SU(2)_G$ theory, since the Higgs branch does not receive
any quantum corrections. This was checked at the level of global
symmetry breaking in
\cite{seiberg/witten:1994.2}, but we can slightly strengthen their result
by explicitly evaluating the metric on the Higgs branch emanating 
from a $U(1)$ theory with $N_e$ massless hypers $Q_{i},\tilde{Q}^i$.

As for the previous case, the Higgs branch is parameterized by a real
parameter up to flavor and gauge rotation. Indeed, by a $SU(N_e)$ rotation
one may choose $Q_i$ along the first flavor; the F--flatness condition
$Q_{i} \tilde{Q}^i = 0$ implies that $\tilde{Q}^i$ has no component along
this direction, so we can use $SU(N_e-1)$ to bring it along the second
flavor:
\BEQ   \pmatrix{ Q_i \cr \tilde{Q}^{i+}} = 
       \pmatrix{ \rho & 0    & 0 & ... & 0 \cr
                 0    & \tilde{\rho}^+ & 0 & ... & 0 }        \EEQ
The D--flatness implies that $\rho=\tilde{\rho}$ up to a $U(1)$ gauge rotation,
and finally one may choose $\rho\in\RR^+$ by a $SU(2)\subset SU(N_e)$
rotation along the two first flavors. We can therefore parameterize the
moduli space by
\BEQ   {\cal M} = \RR^{+} \times {SU(N_e) \over SU(N_e-2) \times U(1)} \EEQ
where the subgroup acts in $SU(N_e)$ through
\footnote{For $N_e=2$ the $U(1)$ is reduced to a $\ZZ_2$ in the center
of $SU(2)$. For $N_e<2$ there is obviously no Higgs branch.}
\BEQ \Omega_{N_e} \stackrel{\equiv} \longrightarrow \Omega_{N_e} .
                          \pmatrix{\II_2 & \cr & \Omega_{N_e-2}} .
			  \pmatrix{e^{i\theta} & \cr & e^{-i\theta/(N_e-2)}}, 
\EEQ
The riemannian structure turns out to be also a warped product:
\BEQ
   ds'^2 = d\rho^2 + \rho^2 \tr \left(\Omega_{N_e}^+ d\Omega_{N_e}
                               \right)_\perp
                 \pmatrix{\II_2 & \cr & } 
                    \left( d\Omega_{N_e}^+ \Omega_{N_e}
                 \right)_{\perp}
\label{metricu1}
\EEQ
where one now projects orthogonally to the $U(1)$ subalgebra.

We can now easily check the duality conjectures on the Higgs branches.
For $N_f$=2, there are two singularities on the Coulomb branch
with two hypermultiplets of
same charge becoming massless at each point. One $\RR^+ \times SU(N_e=2)/\ZZ_2$
Higgs branch emanates from each point, corresponding to the two Higgs branches
$\RR^+ \times SO(4)/SU(2)_G$ of the microscopic $SU(2)_G$ theory. Moreover,
the two metrics (\ref{metricsu2}) and (\ref{metricu1}) coincide.

For $N_f$=3, Seiberg and Witten predict two 
singularities on the Coulomb branch,
one with only 1 charged hyper becoming massless (thereby giving no
Higgs branch), and the other with 4 massless hypers in the spinor of $SO(6)$.
Using the isomorphism $SO(6)=SU(4)/\ZZ_2$, where the $\ZZ_2$ is the square 
of the center $\ZZ_4$ of $SU(4)$
\footnote{This $\ZZ_2$ is actually contained in $SO(2)\times SU(2)_G$,
so that is disappears in the quotient.}, 
one can check that the $SO(2)\times SO(4)$
subgroup of $SO(6)$ translates into the $U(1) \times SU(2) \times SU(2)$ 
subgroup of $SU(4)$ so that we have at the level of differentiable 
manifolds
\BEQ \RR^+ \times {SO(6) \over SO(2) \times SU(2)_G}  \; \equiv \;
     \RR^+ \times {SU(4) \over SU(2) \times U(1) } \EEQ
The metrics can now be seen to coincide, with the peculiarity that the
role of the orthogonal projection on the gauge group on one side is
played by the insertion of $\pmatrix{ \II & \cr &}$ on the other side.

We therefore further check that the global structure as well as the metric
of the Higgs branch is compatible with the singularity structure conjectured
by Seiberg and Witten, and also with the non--renormalization theorem on
the Higgs branch \cite{argyres:1996}
(since those branches coincide not only asymptotically,
but also in the nonperturbative regime).
 
\section{Multicolor Higgs branches}
When the number of colors is increased, the phase structure of the 
theory gets much more involved, in particular since vev's of quarks do
not necessarily completely break the gauge group, so that the gauge
symmetry becomes partially restored on submanifolds of the Higgs branch.
The structure of vacua for $SU(N_c)$ N=2 theories with $N_f$ flavors has
already been carefully analyzed in
\cite{argyres:1996}, with the purpose of
understanding Seiberg's conjectured $N_c\leftrightarrow N_f-N_c$ duality.
In particular, it was noted that the baryonic Higgs branch of a
$SU(N_c)$ theory with $N_f$ flavors could also be interpreted as the
Higgs branch of a $SU(N_f-N_c)\times U(1)^{2N_c-N_f}$ theory with the 
same number 
of flavors and $2N_c-N_f$ color singlets charged under the $U(1)$'s.
Here we shall find another manifestation of this kind of duality,
and prove the exact equivalence (at the level of \hk manifolds)
of the Higgs branches of $U(N_c)$ and $U(N_f-N_c)$ theories with $N_f$
flavors.
As a first hint in this direction, note that the dimension
of the Higgs branch is $N_H = N_f N_c - (N_c^2 -1)=N_c (N_f-N_c)$,
obviously invariant under $N_c \leftrightarrow N_f-N_c$. 
We shall
then comment on the extension of this result to theories with other
gauge groups.

\subsection{$U(N_c)$ N=2 theory with $N_f$ flavors}
The \hk quotient corresponding to a theory $U(N_c)$ with $N_f$ flavors
in the fundamental representation has actually already been worked out
in the mathematical literature
\cite{biquard:1995}, though with very different
motivations than ours. As it turns out, the quotient can be interpreted
geometrically as a cotangent bundle of a complex Grassmannian $G_{N_c,N_f}$.
Now, there is a fairly trivial equivalence of the Grassmannians
$G_{N_c,N_f}$ and $G_{N_f-N_c,N_f}$, and we shall prove that this equivalence
carries over to the \hk structure of their cotangent bundles (that is,
their holomorphic--symplectic structure together with their metric).

Using the same notations as in section \ref{sunc}, the equations describing
the flat directions in presence of a \FI term along the third
direction in $SU(2)_\HH$
\footnote{Whereas in the general case it is not clear how \hk
quotients with rotated \FI terms are related, there is no such problem
here, since a $SU(2)_\HH$ rotation of the \FI terms can be compensated
by a $SU(2)_R$ rotation of the solutions.} read
\BEQA  \tilde{Q} Q &=&0 \\
       Q^+ Q - \tilde{Q} \tilde{Q}^+ &=& 2k \II_{N_c} \EEQA
where $k$ is a fixed real number that can be chosen non--negative.
$Q$ (resp. $\tilde{Q}$) can be seen as a linear endomorphism
$\CC^{N_c} \rightarrow \CC^{N_f}$ (resp. $\CC^{N_f} \rightarrow \CC^{N_c}$ ).

At the level of holomorphic--symplectic manifolds, the \hk quotient
coincides with the quotient of the stable points verifying the first
equation by the action of the complexified gauge group, here $GL(N_c,\CC)$.
The stable points under $GL(N_c,\CC)$ are those for which the matrix $Q$
has maximal rank $N_c$; the invariants under this action are the 
the $N_c$--dimensional vector subspace $P=\im Q\subset \CC^{N_f}$
and the $N_f\times N_f$ meson matrix $M=Q \tilde{Q}$. 
$M$ and $P$ are however constrained by
the F-flatness condition which implies that $P \subset \ker M$. Since
$\im M \subset P$, M is really an endomorphism $\CC^{N_f}/P \rightarrow P$.
The doublet $(P \subset \CC^{N_f}, M:\CC^{N_f}/P \rightarrow P)$ 
has a geometrical interpretation: it defines a point in
the cotangent bundle $T^* G_{N_c,N_f}$ of the complex Grassmannian
$G_{N_c,N_f}$. Indeed the complex Grassmannian $G_{N_c,N_f}$ is the
set of $N_c$--dimensional subspaces $P$ in $\CC^{N_f}$, while tangent
vectors correspond to displacements of $P$, that is linear mappings from
$P$ to its supplementary, $\CC^N_f/P$. $M$ is thus a cotangent
vector at $P\in G_{N_c,N_f}$. Cotangent bundles of complex manifolds
have a canonical holomorphic--symplectic structure,and it can be checked
that it is the same structure as the one obtained by \hk quotient:
\BEQ {\cal M} \stackrel{Hol.Sympl}{\equiv} T^* G_{N_c,N_f} \EEQ
The Grassmannians $G_{N_c,N_f}$ and $G_{N_f-N_c,N_f}$ being isomorphic
as complex manifolds, this is the first hint to the already discussed
duality of the corresponding Higgs branches.

However, we want to show that the isomorphism holds at the level
of \hk manifolds, so we should compare the \kh metric on the 
two quotients. This metric was precisely computed by 
\cite{biquard:1995}, as a
generalization of Calabi's formula 
\cite{calabi:1979} for the \hk on the cotangent bundle
$T\CC P^n$ of the complex projective space, and we shall sketch here their
derivation.
Applying equation (\ref{biquard}), one determines 
the hermitian endomorphism
$g_x\in e^{i{\cal G}^*}$ that takes a point $(Q,\tilde{Q})$ of $M_h$
to $M_0$, such that $g^{-1} Q^+ Q g^{-1} - g \tilde{Q} \tilde{Q}^+ g =
2k 1$ and uses $\chi(g)=det(g)^{2\pi k}$ to describe the \FI term.
As a result,
\BEQ K'(Q,\tilde{Q})= {k\over 2}\ln\det(Q^+ Q)
                      +tr( {1\over 2} \gamma \gamma^+ - {k\over 2}
                       \ln(\gamma\gamma^+) ) 
\label{potun}
\EEQ
where $\gamma=\sqrt{Q^+ Q}g^{-1}$ verifies a biquadratic equation
that yields
\BEQ \gamma\gamma^+ = k(1+\sqrt{1+{1\over k^2}M M^+} ) \EEQ
The first term in (\ref{potun})
is simply the pull back of the \kh metric on $G_{N_c,N_f}$
defined by the scalar product $ds^2=2k tr(XX^+)$ where 
$X:P\rightarrow \CC^{N_f}/P$ is a tangent vector to the Grassmannian.
Indeed, taking $Q=\pmatrix{ \II_{N_c} \cr \cal{Q}}$ one finds
a \kh potential ${k\over2}\ln\det(\II_{N_c} + \cal{Q}\cal{Q^+})$ which
yields upon derivation a generalization of the Fubini--Study metric on
$\CC P^{N_c}$. The second term 
in (\ref{potun}) can be interpreted in terms of the curvature
tensor of $G_{N_c,N_f}$ (cf \cite{biquard:1995}).
In the limit of vanishing \FI term, the \kh potential (\ref{potun}) reduces to
\BEQ K'(Q,\tilde{Q})= {1\over 2} \tr \sqrt{ M M^+}      \EEQ
so that the metric degenerates along the base manifold $G_{N_c,N_f}$.

The isomorphism between the grassmannians $G_{N_c,N_f}$ and $G_{N_f-N_c,N_f}$ 
is obtained by sending a $N_c$-dimensional plane $P$ in $C^{N_f}$ to its
hermitian orthogonal space $Q$ such that $C^{N_f}=P \oplus Q$.
This amounts to take $Q=\pmatrix{ \II_{N_c} \cr \cal{Q}}$ to
$Q=\pmatrix{ \cal{Q}^+ \cr \II_{N_f-N_c}}$.
The cotangent form $M: C^{N_f}/P \equiv Q \rightarrow P$ is then sent to
its hermitian conjugate $M^+: P \equiv C^{N_f}/Q \rightarrow Q$. 
The \kh potential (\ref{potun})
is obviously invariant under this operation, which
proves that the moduli spaces of the $U(N_c)$ and $U(N_f-N_c)$ theories
with $N_f$ flavors are actually the same at the level of \hk manifolds.
This may be relevant for an understanding of Seiberg's duality 
\cite{seiberg:1995} in
$N=1$ SUSY $SU(N_c)$ theory.

\subsection{From $U(N_c)$ to $SU(N_c)$}
We can now fairly easily adapt the previous construction to the case
of a gauge group $SU(N_c)$, where there is no \FI term anymore. 
The flat directions are now given by
\BEQA  \tilde{Q} Q &\propto & \II_{N_c} \\
       Q^+ Q - \tilde{Q} \tilde{Q}^+ &\propto& 2k \II_{N_c} \EEQA
so that the solutions are those of the $U(N_c)$ equations when one
does not impose any value to the \FI terms
\footnote{That is, the moduli space
of $SU(N_c)$ is a fibered over the moduli space of $U(N_c)$, the fiber
corresponding to the three \FI terms together with a $U(1)$ angle
associated to the central $U(1)$ of $U(N_c)$.}.
The complexified gauge group is now $SL(N_c,\CC)$, so that not only is
the subspace $P=\im Q$ preserved, but also the antisymmetric $N_c$--form 
$\varpi$
induced from $\CC^{N_c}$ to $P$ through $Q$:
\BEQ
\varpi(x_1,\dots,x_{N_c}):=det(Q^{-1}x_1,\dots,Q^{-1}x_{N_c}) \hspace{1cm}
			\mbox{for\ } x_i \in \CC^{N_f}
\EEQ
Adding in the $\tilde{Q}$ degrees of freedom and imposing the F--flatness
conditions, we find that the moduli space is actually the cotangent bundle
$T^* G^{V}_{N_c,N_f}$ of the ``complex grassmannian with a volume form''
$G^{V}_{N_c,N_f}$, that is the set of $N_c$--dimensional subspaces $P$
of $\CC^{N_f}$ with any antisymmetric $N_c$--form on $P$.

The metric on this space is simply obtained from the previous construction
by choosing the \FI term $k$ so that $g_x\in G^{\CC}$, that is 
$\det g_x =1$. One then simply finds 
\BEQ 
K'(Q,\tilde{Q})={1\over 2} \tr \sqrt{ k^2 + M M^+ }
\EEQ
with $k$ determined by 
$\det( k \II_{N_f}+\sqrt{k^2 \II_{N_f} + M M^{+}} = \det( Q Q^+)$.

However, the previous duality between $G_{N_c,N_f}$ and $G_{N_f-N_c,N_f}$
has no chance to carry over
to this $SU(N_c)$ case, since there is no way a volume form on $P$
could induce a volume form on its orthogonal.

\subsection{From $SU(N_c)$ to $SO(N_c)$ and $Sp(N_c)$}
In the cases $SO(N_c)$ (resp. $Sp(N_c)$), the gauge group is further reduced,
so that there are more invariants. In particular, from the $Q$ sector
we obtain on the invariant subspace $P$ a bilinear symmetric (resp.
antisymmetric) form, so that we expect the moduli space to be the cotangent
space of the complex Grassmannian ``with a symmetric (resp. antisymmetric)
form''. Once again, given such a form on $P$ there is no way to construct
a form on the orthogonal, and one should not expect the duality
$N_c \leftrightarrow N_f-N_c$ to apply. 

\section{Conclusion}
In this paper, we gave
a self--contained introduction to \hk geometry and studied in detail
some examples of Higgs branches occurring in N=2 SUSY Yang--Mills theories,
complementing the existing literature on the subject.
In particular, we showed how the
\hk structure on the Higgs branch naturally emerges through a
\hk quotient construction, and devoted special attention
to the singularities appearing at points with enhanced symmetry.
While 1--dimensional branches only displayed ADE orbifold singularities,
we explored a higher dimensional example of nontrivial singularity by 
pursuing the work of Refs.\cite{seiberg/witten:1994.2,witten:1996}
on $SU(2)$ and $U(1)$ Higgs branches. We proved the precise equivalence of
the two Higgs branches for special matter content, thereby supporting
both the conjectured singularity structure of the $SU(2)$ theory 
\cite{seiberg/witten:1994.2} and the nonrenormalization theorem on 
Higgs branches \cite{argyres:1996}. Finally, we elaborated on the work
of Ref.\cite{biquard:1995} to prove the invariance of the Higgs branch
of a $U(N_c)\; N_f$ flavors theory under \sd. The connection with Seiberg's
conjectured duality \cite{seiberg:1995} in N=1 is not clear at present,
all the more as this Higgs branch duality does not seem to extend 
to other gauge groups.

\vspace{.1cm}
\begin{flushleft}
  {\it Acknowledgments}
\end{flushleft}
We thank Franck Ferrari, Herv\'e Partouche and Tom Taylor
for useful discussions. B.P. is especially grateful to Olivier Biquard
for many clarifications on the mathematics involved.

\vspace{1cm} 

\centerline{\large\bf Appendix: Riemann tensor of a Warped Product}
\vspace{.4cm}

We consider the warped product $\tilde{M}$ of two Riemannian manifolds
$(M,g)$ and $(M',g')$, of dimension $n$ and $n'$, defined by
\BEQ d\tilde{s}^2 = ds^2 + e^{2\phi(x)} ds'^2  \EEQ
Each vector vector field $\tilde{X}$ on $\tilde{M}$ can be split
along the tangent spaces of $M$ and $M'$ as $\tilde{X}=X+X'$.

The Levi--Civita connection on $\tilde{M}$ corresponding to the above
metric is given by
\BEQA
  \tilde{\nabla}_X Y &=& \nabla_X Y \\
  \tilde{\nabla}_X Y'& =& (X.\phi) Y' \\
  \tilde{\nabla}_{X'} Y &=& (Y.\phi) X' \\
  \tilde{\nabla}_{X'} Y' &=& \nabla'_{X'} Y' - 
                           \tilde{g}(X',Y') \tilde{\nabla}\phi
\EEQA
for which one evaluates the Riemann tensor \\
$R(X,Y)Z:=\nabla_X \nabla_Y Z - \nabla_Y \nabla_X Z - \nabla_{[X,Y]}Z$:
\BEQA
  \tilde{R}(X,Y)Z&=&R(X,Y)Z \\
  \tilde{R}(X,Y)Z'&=&0 \\
  \tilde{R}(X',Y')Z&=&0 \\
  \tilde{R}(X',Y')Z'&=&R'(X',Y')Z'+\| \tilde{\nabla}\phi \|^2
        ( \tilde{g}(X',Z')Y'-\tilde{g}(Y',Z')X') \\
  \tilde{R}(X,Y')Z&=&\left( (X.\phi)(Z.\phi) + 
        \langle \tilde{\nabla}_X d\phi , Z \rangle \right) Y' \\
  \tilde{R}(X,Y')Z'&=& -\tilde{g}(Y',Z') \left( (X.\phi)\tilde{\nabla}\phi
                     + \tilde{\nabla}_X ( \tilde{\nabla} \phi 
                      \right) )
\EEQA
the Ricci tensor $S(X,Y):= tr( X \rightarrow R(X,Y)Z )$:
\BEQA
  \tilde{S}(X,Y)&=&S(X,Y)-n'\left( (X.\phi)(Y.\phi)
                   + \langle \tilde{\nabla}_X d\phi , Y \rangle \right) \\
  \tilde{S}(X',Y)&=&0  \\
  \tilde{S}(X',Y')&=&S'(X',Y')-\tilde{g}(X',Y') 
           \left( \tilde{\Delta}\phi + n'\|\tilde{\nabla}\|^2 \right)
\EEQA
and the scalar curvature
\BEQ \tilde{s} = s+ e^{-2\phi} s'-n'
                 \left( n'\| \tilde{\nabla}\phi\|^2 
                 + \tilde{\Delta}\phi \right)
\EEQ

In the special case where $M$ is 1--dimensional and
 $ \tilde{ds^2} = d\rho^2 + \rho^2 ds'^2 $,
all the components of the Riemann tensor vanish except for
\BEQ
  \tilde{R}(X',Y')Z'=R'(X',Y')Z'+ ( g'(X',Z')Y'-g'(Y',Z')X')
\EEQ
As a check, taking the $n'$--sphere with its canonical
metric for $M'$, the two terms in the last expression
cancel in accordance with $\tilde{M}=\RR^{n'+1}$.
We note that rescaling the metric on the sphere
by a factor distinct from unity turns $\tilde{M}$ into a revolution
cone with a given deficit angle at the apex, and consequently
the cancellation does not take place anymore.


\begin{thebibliography}{10}

\bibitem{seiberg/witten:1994.1}
N. Seiberg and E. Witten, Nucl.Phys. {\bf B426},  19  (1994).

\bibitem{seiberg/witten:1994.2}
N. Seiberg and E. Witten, Nucl.Phys. {\bf B431},  484  (1994).

\bibitem{Barbieri:1982}
R. Barbieri, S. Ferrara, L. Maiani, F. Palumbo and
C. A. Savoy, Phys.Lett. {\bf 115B},  212  (1982).

\bibitem{argyres:1996}
P.~C. Argyres, M.~R. Plesser, and N. Seiberg, hep-th/{\bf 9603042}.

\bibitem{biquard:1995}
O. Biquard and P. Gauduchon,  in {\em Proceedings on Geometry and Physics
  (Aarhus)}, edited by J. Andersen, J. Dupont, H. Pedersen, and A. Swann
  (Birkhauser, Basel, 1995), to appear.

\bibitem{alvarez-gaume:1981}
L. Alvarez-Gaum\'e and D.~Z. Freedman, 
Commun.Math.Phys. {\bf 80},  443  (1981).

\bibitem{hitchin:1987}
N.J. Hitchin, A. Karlhede, U. Lindstr{\"o}m, and M. Ro{\v c}ek, 
Commun.\-Math\-
.Phys. {\bf 108},  535  (1987).

\bibitem{galicki:1987}
K. Galicki, Commun.Math.Phys. {\bf 108},  117  (1987).

\bibitem{hitchin:1993}
N. Hitchin, Seminaire Bourbaki {\bf 748},  1  (1991-2).

\bibitem{billo:1994}
M. Billo and P. Fr\'e, hep-th/{\bf 9411183}.

\bibitem{fre:1995}
P. Fr\'e, hep-th/{\bf 9512043}.

\bibitem{kobayashi:1963}
S. Kobayashi and K. Nomizu, {\em Foundations of Differential 
  Geometry vol.I and
  II} (Interscience publishers, J. Wiley and Sons, New York, 1963).

\bibitem{nakahara:1990}
M. Nakahara, {\em Geometry,Topology and Physics} (Inst. of Physics Publishing,
  Bristol, Philadelphia, 1990).

\bibitem{andrianopoli:1996}
L. Andrianopoli, M. Bertolini, A. Ceresole, R. D'Auria, S. Ferrara, P. Fr{\'e} 
and T. Magri, hep-th/{\bf 9605032}.

\bibitem{antoniadis:1996.1}
I. Antoniadis, H. Partouche, and T.~R. Taylor, Phys.Lett. {\bf B372},  83
  (1996).

\bibitem{antoniadis:1996.2}
I. Antoniadis and T.~R. Taylor, hep-th/{\bf 9604062}.

\bibitem{marsden:1982}
J. Marsden and A. Weinstein, Rep.Math.Phys. {\bf 5},  121  (1974).

\bibitem{witten:1993}
E. Witten, Nucl.Phys. {\bf B403},  159  (1993).

\bibitem{kronheimer:1989}
P.~B. Kronheimer, J. Diff. Geom. {\bf 29},  665  (1989).

\bibitem{eguchi/hanson:1978}
T. Eguchi and A.J. Hanson, Phys.Lett. {\bf 74B},  249  (1978).

\bibitem{eguchi/hanson:1979}
T. Eguchi and A.J. Hanson, Ann.Phys. {\bf 120},  82  (1979).

\bibitem{hitchin:1994}
N.~J. Hitchin, Math.Proc.Camb.Phil.Soc {\bf 85},  465  (1979).

\bibitem{duistermatt:1982}
J.J. Duistermaat and G.J. Heckman, Invent.Math. {\bf 69},  259  (1982).

\bibitem{anselmi:1994}
D. Anselmi, M. Bill\`o, P. Fr\', L. Girardello and A. Zaffaroni, 
Int.\-J.\-Mod.\-Phys. {\bf A9},  3007  (1994).

\bibitem{nakajima:1994}
H. Nakajima, Duke Math.J. {\bf 76},  365  (1994).

\bibitem{douglas:1996}
M.~R. Douglas and G. Moore, hep-th/{\bf 9603167}.

\bibitem{witten:1996}
E. Witten, Nucl.Phys. {\bf B460},  541  (1996).

\bibitem{wess/bagger:1992}
J. Wess and J. Bagger, {\em Supersymmetry and Supergravity}, {\em Princeton
  Series in Physics} (Princeton University Press, Princeton, New Jersey, 1992),
  second edition.

\bibitem{besse:1986}
A.~L. Besse, {\em Einstein Manifolds} (Springer-Verlag, Berlin Heidelberg,
  1986).

\bibitem{atiyah:1978}
M.F. Atiyah, V.G. Drinfeld, N.J. Hitchin, and Y.I. Manin, Phys.Lett. {\bf 65A},
  185  (1978).

\bibitem{calabi:1979}
E. Calabi, Ann.\-Ec.\-Norm.\-Sup. {\bf 12},  269  (1979).

\bibitem{seiberg:1995}
N. Seiberg, Nucl.Phys. {\bf B435},  129  (1995).

\end{thebibliography}
 

\end{document}